\documentclass[11pt]{article}

 \newcommand{\be}{\begin{equation}}
 \newcommand{\ee}{\end{equation}}
 \newcommand{\ba}{\begin{eqnarray}}
 \newcommand{\ea}{\end{eqnarray}}
 \newcommand{\bl}{\begin{equation}\begin{array}{ll}}
 \newcommand{\el}{\end{array}\end{equation}}
 \newcommand{\bll}{\begin{equation}\begin{array}{lll}}
 \newcommand{\bdm}{\begin{displaymath}}
 \newcommand{\edm}{\end{displaymath}}
 \def\p{\partial}
 \def\f{\varphi}
 \def\ve{\varepsilon}
 \def\ra{\rightarrow}
\def\lim{\rightarrow}
\def\half{\frac{1}{2}}
\def\quart{\frac{1}{4}}

\begin{document}
\raggedbottom

\title{Integrable Low Dimensional Theories Describing High Dimensional
Branes, Black Holes and Cosmologies}

\author{A.T.~Filippov\thanks{filippov@thsun1.jinr.ru}\\
{\small \it  Joint Institute for Nuclear Research, Dubna, Moscow region RU-142980}
\and
V.~de~Alfaro\thanks{vda@to.infn.it}\\
{\small \it DFTT University of Turin, INFN section of Turin, Turin, I-10125}}


\maketitle

\begin{abstract}

The reduction of higher dimensional supergravities to low dimensional dilaton
gravity theories is outlined. Then a recently proposed new class of integrable
theories of 0+1 and 1+1 dimensional dilaton gravity coupled to any number of
scalar fields is described in more detail.
These models are reducible  to systems of independent Liouville
equations whose solutions are not independent because they must
satisfy the energy and momentum constraints.
The constraints are solved, thus giving the explicit analytic
solution of the theory in terms of arbitrary chiral fields.
In particular, these integrable theories describe spherically
symmetric black holes and branes of higher dimensional
supergravity theories as well as superstring motivated
cosmological models. Note that the reader is strongly recommended to
have a look into the transparencies of the lectures on which this text
is based (http://www.phy.bg.ac.yu/mphys2/)\footnote{In addition, we recommend
the reader to consult our
papers \cite{VDA}, \cite{A1}, \cite {atfm} and the reviews \cite{S}, \cite{St},
\cite{Mo}, \cite{Kummer}  for more details and references.}.
\end{abstract}

\section{Introduction}

The low dimensional field theories were usually considered as
toy models for the `real' four dimensional theory. Even the fact that
some solutions of the higher dimensional theories may be consistently
considered as low dimensional field theories or even one dimensional
dynamical systems did not change the general attitude to low dimensional
theories. The situation began to change with the development of string theory
in which the connection between high and low dimensions is obvious and
profound. Especially interesting are low dimensional dilaton gravity
theories which are capable to describe some physically interesting
phenomena in higher dimensions.

In fact, in the last decade 1+1 and 0+1 dimensional dilaton gravity
coupled to scalar matter fields proved to be a reliable
model for higher dimensional black holes and string inspired
cosmologies. The connection between high and low dimensions has
been demonstrated in different contexts of gravity and string
theory - symmetry reduction, compactification, holographic
principle, AdS/CFT correspondence, duality. For spherically symmetric
configurations the description of static black holes, branes and of
cosmological solutions even simplifies to 0+1 dimensional dilaton gravity
- matter models, which in many interesting cases are explicitly
analytically integrable (see e.g. \cite{CGHS} - \cite{Iv}
and references therein).

However, generally they are not integrable. For example,
spherical black holes coupled to Abelian gauge fields are usually
described by integrable 0+1 dimensional models, while the addition of the
cosmological constant term destroys integrability. In 1+1 dimension,
pure dilaton gravity is integrable but the coupling to scalar matter
fields usually destroys integrability. The one very well studied exception
is the CGHS model. In \cite{A1} a more general
integrable model of dilaton gravity
coupled to matter, which incorporates as limiting
cases the CGHS and other known integrable models  was proposed.
It reduces to two Liouville
equations, whose solutions should satisfy two constraints.
Because the general analytic solution of the constraints had been
not found at that time, the model of Ref.\cite{A1} received little
attention and was not studied in detail\footnote{
Some applications of this model were discussed in \cite{Nav}. Note also that
some recent cosmological models use potentials similar to those introduced
in \cite{A1}. We will consider such applications in a separate paper.}.

Recently, one of the authors has proposed a class of more
general integrable dilaton gravity models in dimension 1+1, which
are reducible to $N$ Liouville equations (a brief summary is
published in Ref.\cite{A2}). For these models the general
analytic solution of the constraints has been found.
It was demonstrated that the $N$-Liouville models are
closely related to physically interesting solutions of higher
dimensional supergravity theories describing the low energy limit of
superstring theories. These 1+1 dimensional $N$-Liouville theories describe
the solutions of higher dimensional theories in some approximation.
On the other hand,
their reduction to dimensions 1+0 (cosmological) and 0+1 (static
black holes) give the exact solution of higher dimensional theories.

Static black holes and cosmological models are described by one
dimensional solutions of the 1+1 dimensional theories. In the
standard approach the deep connection between black holes and cosmologies
 is not transparent and is usually ignored (even the precise
relation between the dimensional reductions used by `cosmologists' and
by `black holes investigators' is not quite obvious).
We thus start from the 1+1 dimensional formulation to get a unified
description of these two objects. A characteristic feature of the
static solutions of the models derived from string theory
is the existence of horizons with nontrivial scalar field
distributions (what must be characteristic features of string cosmologies
is as yet a much discussed problem).

It is well known that in the Einstein - Maxwell theories
minimally coupled to scalar fields the spherical static horizons disappear
if the scalar fields have a nontrivial space distribution (this is the so-called
 `no-hair theorem'). In Ref. \cite{A1}, a local version of the no-hair theorem
 (we call it the `no horizon' theorem) was formulated and proved. It states
that, under certain conditions, there exists no static solution with a
horizon in a class of 0+1 dimensional dilaton gravity
theories coupled to scalar matter (the important requirement
is that the scalar fields vary in space and are finite on the horizon).
The theorem is local, and does not require any boundary conditions at
infinity.

However, the `no horizon' theorem is not true (as is known in
several examples) for Einstein - Yang-Mills theories \cite{M1}
as well as for solutions of higher
dimensional supergravities, see e.g. \cite{Kiem}.
In all these cases the static solutions of
higher dimensional theories may be constructed by using the 1+1 or
0+1 dimensional dilaton gravity coupled to
matter. In the integrable models we discuss here the solutions
with horizons are completely identified and described in very
simple terms. One may also consider the global properties of the
solutions with or without horizons but we will not discuss this
subject here.

In Section~2 we briefly demonstrate that spherically
symmetric black holes and branes of higher dimensional
supergravity theories, as well as superstring motivated
cosmological models, may be described in terms of 0+1 and
1+1 dimensional dilaton gravity theories. In Sections~3,~4
a new class of integrable theories of 0+1 and 1+1 dimensional dilaton
gravity coupled to any number of scalar fields is discussed.
In Section~5 we outline possible applications of the integrable
models and some unsolved problems. In Appendix we present a derivation
of some results and some useful formulas. Note that the models with
nonlinear coupling of gauge fields to dilaton gravity were not
considered in the literature in full generality and the construction of
the effective potential for them is, to the best of our knowledge,
a new result (this result was mentioned in \cite{A3} and in the report
of one of the authors (A.T.F.) to the Third Sakharov Conference
but its proof was not published).

\section{High dimensional dilaton gravity}

We first write the higher dimensional theories which, under
dimensional reductions, produce special examples of integrable
theories introduced in \cite{A1} and \cite{A2}. They all come from the low
energy limit of the superstring theories, which is described by
10 dimensional supergravities\footnote{
One may also start with the 11 dimensional supergravity, which is believed
to be related to the so called `M - theory', and reduce to 10 dimension
by compactifying one dimension. Note also that here we are not attempting to consider
compactifications of the most general supergravities. We only demonstrate the
main features of the connection between low dimensional and high dimensional
theories}.
The bosonic part of the 10
dimensional supergravities of type II (corresponding to the type
II superstrings) may be written as
 \be
 {\cal L}^{(10)} =
{\cal L}_{NS-NS}^{(10)} \, + \, {\cal L}_{RR}^{(10)} \, , \label{eq:1}
\ee
In this brief discussion it is sufficient to consider the first
Lagrangian\footnote{
The second one gives similar 1+1 dimensional theories.
Eventually, all bosonic and fermionic parts of the higher dimensional
Lagrangians give in 1+1 dimension dilaton gravity coupled to scalar matter
fields.}:
\be
{\cal L}_{NS-NS}^{(10)} \, = \sqrt{-g^{(10)}} e^{-2\phi_s} \biggl[ R^{(10)} +
4(\nabla \phi_s)^2 - {1\over 12} H_3^2\biggr] , \label{eq:2}
\ee
Here $\phi_s$ is the dilaton, related to the string coupling
constant; $H_3 = dB_2$ is a 3-form; $g^{(10)}$ and $R^{(10)}$ are the
10 dimensional metric and scalar curvature respectively.

There are many ways to reduce high dimensional theories to low dimensions
1+1 and 0+1. We only mention here those that may lead to integrable
theories. First, one may compactify the $d$ dimensional theory on
a $p$ dimensional torus $T^p$ (or on several tori, including the circle $S^1$)
using the Kaluza - Mandel - Fock - Klein  (KMFK) mechanism\footnote{
It is usually called the Kaluza - Klein (KK) mechanism but this is not justified
historically. Actually, the Russian theorists George Mandel and Vladimir Fock have written
their papers, in which they generalized the Kaluza theory, even somewhat earlier
than Oscar Klein and published them in the same journal. We hope to redress this historical
injustice in a separate publication}.
This introduces $p$ Abelian gauge fields and at least $p$ scalar fields. Antisymmetric
tensor fields ($n$-forms), which may be present in the high dimensional theory,
will produce lower-rank forms and, eventually, other scalar fields. Thus we get
a theory of gravity coupled to matter fields (scalars, Abelian gauge fields and,
possibly, higher-rank forms) in a $d$ dimensional space, $d = D -p$. The next step
is to reduce further its dimension by using some symmetry of the $d$ dimensional theory.
The most typical one is the spherical symmetry (the axial symmetry leads to much
more complex low dimensional theories and is not considered here). This step produces
a 1+1 dimensional dilaton gravity theory coupled to scalar and gauge fields.
The simplest example is the spherical reduction of the $d$ dimensional Einstein -
Maxwell theory - the resulting 1+1 dimensional dilaton gravity is actually equivalent
to a 0+1 dimensional theory.

The 1+1 dimensional dilaton gravity theories so derived may describe static black holes
(static solutions), spherically symmetric evolution of the black holes (collapse of
matter) and of the universe (expansion of the universe). In this sense, the flat space
homogeneous cosmological models and static black holes may be regarded as the 1+0 and 0+1
dimensional reductions of the 1+1 dimensional theory and they can be connected in the frame
of the 1+1 dimensional model.

Note that the final step in the chain of dimensional
reductions in cosmology is usually somewhat different from that in black hole physics.
Cosmological models are normally obtained by reducing the $d$ dimensional theory
directly to dimension 1+0. Indeed, isotropy and homogeneity of the universe require
a higher symmetry than the spherical one - the whole space should have constant
curvature $k$, which may be equal to zero or $\pm 1$. These cosmologies can be selected
from the set of the 1+0 dimensional solutions of the 1+1 dimensional theory by choosing
a proper dimensional reduction of the metric and of the dilaton. We will not go into
a detailed description of dimensional reductions, referring the reader to an instructive
example in \cite{BMG}, \cite{VDA}, \cite{Kiem} and to reviews \cite{St} - \cite{Iv}.
Instead we give a simplified typical chain of dimensional reductions leading to
simple and important two dimensional and one dimensional dilaton gravity models.

Reducing to $d$ dimensions by different sorts of dimensional
reduction (KMFK, compactification on tori, etc.) we obtain an
effective Lagrangian ${\cal L}^{(d)}$. For our purposes it is sufficient to consider
the following expression
\ba
{\cal L}^{(d)} = \sqrt{-g^{(d)}} e^{-2\phi_d} \biggl( R^{(d)} + 4(\nabla
\phi_d)^2 - {1\over 2} (\nabla \psi)^2 -  \nonumber \\
- X_0 - X_1 (\nabla
\sigma)^2 - X_2 F_2^2 \biggr). \label{eq:3}
\ea
Here $\phi_d$ is a new dilaton, $F_2$ is a 2-form (an Abelian
gauge field); $X_a$ are functions of
$\phi_d$ and $\psi$. Actually, the Lagrangian should depend on
several $F_2$-fields, several $\psi$-fields, and may depend on several
$\sigma$-fields  as well as on higher - rank forms. However,
after further reduction to two dimension only 2-forms and scalar
fields will survive (in fact, the 2-forms can also be excluded by writing
an effective potential depending on electric or magnetic charges, see below).

We further reduce the $d$ dimensional theory to dimension 1+1 by spherical
symmetry. Before and after doing so one may transform this Lagrangian by
the Weyl conformal transformation, $g_{\mu\nu} \Rightarrow \tilde{g}_{\mu\nu}
\equiv \Omega^2 g_{\mu\nu}$,
where $\Omega$ depends on the dilaton. Expressing $R$ in terms of the
new metric,
\be
R = \Omega^2 \biggl[ \tilde{R} + 2(d-1) {\tilde{\nabla}}^2 \ln{\Omega} -
(d-1)(d-2) (\tilde{\nabla} \ln{\Omega})^2 \biggr] \, , \label{eq:3a}
\ee
one can easily find the new expression for the Lagrangian. For $d > 2$ one can
cancel the multiplier $e^{-2\phi_d}$ by choosing an appropriate function
$\Omega(\phi_d)$ and thus write the Lagrangian in the so called Einstein frame
(as distinct from the string frame expressions above). In dimension $d=2$
it is impossible to remove the dilaton multiplier but, instead, one can remove
the dilaton gradient term.

Now consider the spherically symmetric solutions of the $d$ dimensional theory (3).
Usually, it is more convenient to remove the dilaton factor by a Weyl transformation
and rewrite the action (3) in the Einstein frame,
\ba
{\cal L}^{(d)}_E = \sqrt{-g^{(d)}} \biggl[ R^{(d)} - \half (\nabla \chi)^2
- \half (\nabla \psi)^2 -
\nonumber \\
- X_0 e^{a_0 \chi} - X_1 e^{a_1 \chi } (\nabla\sigma)^2 -
 X_2 e^{a_2 \chi } F_2^2 \biggr], \label{eq:3b}
\ea
where $\phi_d \equiv \chi$ and $a_k$ are known constants depending on $d$.
Then we parameterize the spherically symmetric metric by
the general 1+1 dimensional metric $g_{ij}$ and the dilaton $\f$ ($\nu \equiv 1/n$,
$n \equiv d-2$),
\be
ds^2=g_{ij}\, dx^i\, dx^j \,+\, {\f}^{2\nu} \,
d\Omega_{(d-2)}^2 \, , \label{eq:4}
\ee
introduce appropriate spherical symmetry conditions for the
fields, which from now on will be functions of the variables
$x_0$ and $x_1$ ($t$ and $r$), and integrate out the other (angular)
variables from the $d$ dimensional action.

Applying, in addition, the Weyl transformation that removes the dilaton gradient
term we thus obtain the effective 1+1 dimensional action
\ba
{\cal L}= \sqrt{-g} \biggl[ \f R + n(n-1) \f^{-\nu} -
X_0 e^{a_0 \chi } \f^{\nu} -
X_2 e^{a_2 \chi} \f^{2-\nu} F^2 -
\nonumber \\
 - \half \f \biggl( (\nabla \chi )^2
+ (\nabla \psi)^2 + 2 X_1 e^{a_1 \chi } (\nabla \sigma)^2
\,\biggr) \, \biggr] . \label{eq:5}
\ea
Here $\f$ is the 2-dilaton field that is often denoted by $e^{-2\phi}$;
the scalar fields $\psi$ may have different origins -- they may be former
dilaton fields, KMFK scalar fields, reduced $p$-forms, etc. The functions $X_k$ (we call
them potentials) depend on the scalar fields $\chi$ and $\psi$, which
from now on will be called scalar matter fields. Also the field $\sigma$ may be
regarded as a matter field but it plays a special role that will be discussed later.
Notice that the potentials are positive and that $n(n-1)$
is positive or zero\footnote{
This term is the curvature of the
$n$ dimensional sphere whose metric is given by the second term in (5).
If, instead of the spherical symmetry, we used
a pseudo-spherical one, the sign would be negative. If the $n$ dimensional
subspace is flat this term will be absent.}.

For dimensionally reduced supergravity theories one can often find a
parameterization of the fields in which the potentials are exponentials
of the matter fields while the kinetic (gradient) terms have the above simple structure.
These 1+1 dimensional theories may have an integrable one dimensional
sector describing static (0+1) or cosmological (1+0) solutions of the higher dimensional
theories. The 1+1 dimensional theories obtained by dimensional reductions are
usually not integrable but may often be approximated by explicitly analytically integrable
1+1 dimensional theories.

As it was mentioned above, the cosmological models are usually obtained from higher
dimensional theories by a different dimensional reduction. To describe the homogeneous
and isotropic universe one supposes that the metric may be written in the form
\be
ds^2 = -e^{2\nu(t)} dt^2 + e^{2\mu(t)} d\Omega^2_{(d-1)}(k) \, , \label{eq:5a}
\ee
where $k=0$ for the flat space and $k=\pm 1$ for the space of constant positive
(negative) curvature.

Now, in cosmological models somewhat different reductions of
the fields are of interest because the terms generated by the higher rank forms
(characteristic of string theories) are believed to be of interest. However, after
reducing to one dimension, also the higher rank forms give scalar fields either
$\psi$ or $\sigma$ type. For example, in a typical reduction of the type IIA
10 dimensional supergravity to dimension 4 (compactification on an isotropic
six dimensional torus $T^6$) and then to 1+0 dimensional dilaton gravity
(see e.g. \cite{Bill}), the 3-form produces in the one dimensional theory a $\sigma$
term while the 4-form generates an $X_0$-type potential. The  cosmologies so obtained
are in general not integrable.


\section{1+1 dimensional dilaton gravity}

Now let us consider a general 1+1 dimensional dilaton gravity
coupled to Abelian gauge fields $F^{(a)}_{ij}$ and to scalar
fields $\psi_n$. The general Lagrangian can be written as
\ba
{\cal L} = \sqrt{-g} \, [ \, U(\f) R(g) +
V(\f) + W(\f)(\nabla \f)^2 + \nonumber \\
+ X (\f,\psi, F_{(1)}^2, ..., F_{(A)}^2) + Y(\f,\psi)+ \sum_n Z_n(\f,\psi)
(\nabla \psi_n)^2 \, ] \, . \label{eq:6}
\ea
Here $g_{ij}$ is a generic 1+1 dimensional metric with signature (-1,1) and
$R$ is the Ricci curvature; $U(\f)$, $V(\f)$, $W(\f)$ are
arbitrary functions of the dilaton field; $X$,
$Y$ and $Z_n$ are arbitrary functions of the
dilaton field and of $(N-2)$ scalar fields $\psi_n$ ($Z_n < 0)$;
$X$ also depends on $A$ Abelian gauge fields
 $F_{(a)ij} \equiv F^{(a)}_{ij}$,
 $F_{(a)}^2 \equiv g^{ii'} g^{jj'} F^{(a)}_{ij} F^{(a)}_{i'j'}$.
Notice that in dimensionally reduced theories (see (\ref{eq:5}))
both the scalar fields and the Abelian gauge fields are
non-minimally coupled to the dilaton.

The equations of the theory (\ref{eq:6})
can be solved for arbitrary potentials $U$, $V$, $W$ and $X$ if
$\p_{\psi}X \equiv 0$
(for the simplest explicit solution in case of $X$ linear in $F^2$
see e.g \cite{A1} and references therein as well as the recent review \cite{Kummer}).
Actually, only $V(\phi)$ and $X$ are really arbitrary functions.
Moreover, for general potentials $X(\phi, \psi, F^2)$ one may solve the equations for
$F^{(a)}_{ij}$ and then construct the effective action (see Appendix)
\be
{\cal L}_{\rm eff} =\sqrt{-g}\left[ \f R(g) + V_{\rm eff}(\f,\psi) +
\sum_n Z_n(\f,\psi)\, g^{ij} {\p}_i \psi_n {\p}_j \psi_n \right]\,.
\label{eq:7}
\ee
Here the effective potential $V_{\rm eff}$ (below we omit the subscript) depends
also on charges introduced by solving the equations for the Abelian
fields. Note also that we use a Weyl transformation to exclude the kinetic
term for the dilaton and also choose the simplest, linear parameterization
for $U(\f)$\footnote{
If $U^{\prime}(\f)$ has zeroes, this parameterization, as well as
the more popular exponential one, $U = e^{\lambda \phi}$, is valid only between
two consecutive zeroes.}.

If  the effective potential does not depend on $\psi$, one can find the general
solution for the matter vacuum when all $\psi$ are constant. In this case the
equation of motion actually reduce to those of the pure dilaton gravity not
coupled to scalar matter. Few 1+1 dimensional models are integrable. The best studied
ones are the CGHS and JT models.  They were essentially generalized in \cite{A1}.
In all these models the $Z$-potentials are constant (so called minimal coupling).
The only integrable class of models with non minimal
coupling to scalar fields (with some special functions $Z_n(\phi)$)
was proposed in \cite{FI}.


Now we introduce a more general class of integrable
1+1 dimensional dilaton gravity models with minimal coupling to scalar fields.
They are defined by the Lagrangian (\ref{eq:7})
with the following potentials:
\be
Z_n = -1; \;\;\;\;  |f|V = \sum_{n=1}^N 2g_n e^{q_n} \, . \, \label{eq:8}
\ee
Here $f$ is the light cone metric, $ds^2 = -4f(u, v) \, du \, dv$, and
\be
q_n \equiv F + a_n \phi + \sum_{m=3}^{N} \psi_m a_{mn} \equiv
\sum_{m=1}^{N} \psi_m a_{mn} \, , \label{eq:9}
\ee
where $\psi_1 + \psi_2 \equiv \ln{|f|} \equiv F$ ($f \equiv \varepsilon e^F$)
and $\psi_1 - \psi_2 \equiv \phi$.-
Now, varying the Lagrangian (\ref{eq:7}) in $(N-2)$ scalar fields, dilaton and in
$g_{ij}$ and then passing to the light cone metric we find $N$ equations of motion
for $N$ functions $\psi_n$,
\be
\epsilon_n \p_u \p_v \psi_n  =  \sum_{m=1}^{N} \varepsilon g_me^{q_m} a_{mn} \, ; \,\,\,\,
\epsilon_1 = -1, \,\,\, \epsilon_n = +1, \,\, {\rm if} \,\, n \geq 2 \, ,
\label{eq:10}
\ee
and two constraints,
\be
C_i \equiv f {\p}_i ({\p}_i \phi /f) + \sum_{n=3}^N ({\p}_i \psi_n)^2 = 0 ,
\,\,\, i = (u, v) .
\,\, \label{eq:11}
\ee


For arbitrary coefficients $a_{mn}$ the equation of motion are not integrable.
However, if the $N$-component vectors $v_n \equiv (a_{mn})$ are
pseudo - orthogonal, the equations of
motion can be reduced to $N$ Liouville equations for $q_n$,
\be
{\p}_u {\p}_v q_n - {\tilde{g}}_n e^{q_n} =0 , \label{eq:12}
\ee
where ${\tilde{g}}_n = \ve \lambda_n g_n$, $\lambda_n = \sum \epsilon_m a_{mn}^2$,
and $\ve \equiv |f|/f$ (note that the equations for $q_n$ depend on $\epsilon_n$
only implicitly, through the normalization factor $\lambda_n$).

The most important fact is that the constraints can be explicitly solved.
By writing the solutions of the Liouville equations in the form suggested
by the conformal field theory,
\be
e^{-q_n /2} = a_n(u) b_n(v) + \bar{a}_n(u) \bar{b}_n(v) ,
\label{eq:13}
\ee
where $\bar{a}$ and $\bar{b}$ can be expressed in terms of $a$ and $b$, i.e.
\be
e^{-q_n /2} = a_n(u) b_n(v) \biggl[ 1 - \half {\tilde{g}}_n
\int {du \over a_n^2(u)} \int {du \over b_n^2(v)} \biggr] \, ,
\ee
we may rewrite the constraints in the form
\be
C_u = \sum_{n=1}^N a_n^{\prime \prime} (u) [\lambda_n a_n(u)]^{-1} \, , \,\,\,\,
C_v = \sum_{n=1}^N b_n^{\prime \prime} (v) [\lambda_n b_n(v)]^{-1} \, .
\label{eq:14}
\ee
Using the fact that the norms $\lambda_n$ satisfy the constraint
$\sum_{n=1}^N \lambda_n^{-1} = 0$ (this is a consequence of the pseudo-orthogonality
conditions) we can solve these constraints. The solution has the
following form:
\be
{{a_n^{\prime} (u)} \over {a_n(u)}} = \alpha_n (u) -  {{\sum_{n=2}^N {\lambda_n}^{-1}
(\alpha_n^{\prime} + \alpha_n^2)} \over {2\sum_{n=2}^N {\lambda_n}^{-1} \alpha_n}} ,
\label{eq:15}
\ee
where $\alpha_1 = 0$ and the other $\alpha_n$ are arbitrary functions of $u$.
The ratios $b_n^{\prime}(v) / b_n(v)$ are expressed by the same equation in terms
of functions $\beta_n(v)$.

By integrating the first order differential equations for $a_n(u)$ and $b_n(v)$
we thus find the general solution of the $N$-Liouville dilaton gravity in terms
of $(2N-2)$ arbitrary chiral fields $\alpha_n(u)$ and $\beta_n(v)$.
With proper conditions of convergence one may use this solution also for $N=\infty$.


\section{Integrable 0+1 dimensional dilaton gravity coupled to matter}

The dimensional reduction from 1+1 to 0+1 dimensions in the light cone
coordinates $(u, v)$ is very simple.
 If we suppose that $\f=\f(\tau)$, $\psi_n =\psi_n(\tau)$ where
 $\tau=a(u)+b(v)$, we find from the 1+1 dimensional equations of motion that
 \be
 f(u,v) = \mp h(\tau) \, a'(u) \, b'(v)
 \label{eq:16}
 \ee
and thus
\be
 ds^2 = -4f(u,v) \, du \, dv = \pm 4h(\tau) da db.
\label{eq:16a}
\ee
Defining two dimensional space and time coordinates $r = a \pm b$ and
$t = a \mp b$ we find that
\be
ds^2 = h(\tau) (dt^2 - dr^2) , \,\,\,\, {\rm where} \,\,\,\,\,
\tau = r \,\, {\rm or} \,\,
\tau = t ,
\label{eq:17}
\ee
and thus the reduced solution may be the static or the cosmological
one\footnote{
Of course, in 2d theories this distinction is not very important.
However, when we know the higher dimensional theory from which our 2d dilaton
gravity originated, we can reconstruct the higher dimensional metric and thus
find the higher dimensional interpretation of our solutions. In the remaining part
of these lectures we do not introduce $r$ and $t$, take in (\ref{eq:16}) the upper sign
and usually call all one dimensional solutions static.}.

However, this is not the most general way for obtaining 0+1 or 1+0 dimensional
theories from higher dimensional ones. Not all possible reductions can be derived
by this simple dimensional reduction of the 1+1 dimensional gravity. For example,
to derive the cosmological solutions corresponding to the reductions (\ref{eq:5a})
one should use a more complex dimensional reduction of the 1+1 dimensional dilaton
gravity, which will be discussed elsewhere.

It is not difficult to show that the 0+1 dimensional equations are described
by the Lagrangian ($\ln{|h|} = F$, $\ve = \pm$) \cite{A1}:
\be
{\cal L} =-{1\over l} \biggl( \dot \f \dot F +\sum_n
Z_n(\f, \psi) \dot \psi_n^2 \biggr) +l\, \ve e^F \, V(\f , \psi) ,
\label{eq:18}
\ee
where $l(t)$ is the Lagrange multiplier (related to the general metric $g_{ij}$).

Now, the two-dimensionally integrable $N$-Liouville theories presented above are also
integrable in 0+1 dimension. Moreover, as we can solve the Cauchy problem in dimension
1+1 we can study the evolution of the initial configurations to stable static solutions,
e.g. black holes, which are special solutions of the 0+1 dimensional reduction.
However, the reduced theories can be explicitly solved for much more general potentials
$Z_n$ and $V$.

Suppose that for $(N-2)$ scalar fields $\psi_n$ ($n = 3,...,N$) the
ratios of the $Z$-potentials are constant so that we can write
$Z_n = -\gamma_n /\phi^{\prime}(\f)$ (in eq.~(\ref{eq:5}) these are the fields $\chi$ and
$\psi$ and $\phi = \ln \f$). Suppose that all the potentials $Z_n$ and $V$ be
independent of the scalar fields $\psi_{N+k}$ with $k = 1,...,K$ (in eq.~(\ref{eq:5})
this is the field $\sigma$). Then, we first remove the factor $\phi^{\prime}(\f)$
by defining the new Lagrange multiplier $\bar{l} = l(\tau) \phi^{\prime}(\f)$ and absorb the
factors $\gamma_n > 0$ in the corresponding scalar fields. In this way we may introduce
the new dynamical variable $\phi$ instead of $\f$. Now we can solve the equations for
the $\sigma$-fields and construct the effective Lagrangian\footnote{
It is better to do this in the Hamiltonian formalism but space limitations force us
to omit details of our derivations.}.
We thus may arrive at the effective Lagrangian
\be
{\cal L}_{\rm eff} =-{1\over \bar{l} }\biggl[ \dot\phi \dot F - \sum_{n=3}^N
 \dot \psi_n^2 \biggr] + \bar{l} \biggl[ \ve e^F  V_{\rm eff}(\phi, \psi) +
 V_{\sigma} (\phi, \psi) \biggr] .
\label{eq:19}
\ee
Here $V_{\rm eff} = V/\phi^{\prime}(\f)$ and $V_{\sigma} =
\sum_k C_k^2 /Z_{N+k} \, \phi^{\prime}(\f)$ where $\f$ must be expressed in terms of
$\phi$.
If the original potentials in eq.~(\ref{eq:18}) are such that
$(Z_{N+k} \, \phi^{\prime}(\f))^{-1}$ and
$V / \phi^{\prime}$ can be expressed in terms of sums of exponentials of linear combinations
of the fields $\phi$ and $\psi$, then there is a chance that the 0+1 dimensional
theory can be reduced to Liouville or Toda equations (the Toda case is possible
only if $V_{\sigma} \neq 0$).

The pure Liouville case was introduced in \cite{A2}
and is described by the Lagrangian (in notations of eq.~(\ref{eq:9}))
\be
{\cal L} = {1\over l} \bigl( - \dot \psi_1^2 +
 \sum_{n=2}^N \dot \psi_n^2 \bigr) +l \sum_{n=1}^N 2g_n
e^{q_n}.
\label{eq:20}
\ee
If the $a_{mn}$ satisfy our pseudo orthogonality conditions, the equations of motion are
reduced to $N$ independent one dimensional Liouville equations whose solutions
have to satisfy the energy constraint. The solutions are expressed in terms of
elementary exponentials (for simplicity, we write the solution in the gauge
$l(\tau) \equiv 1$ but all the results are actually gauge invariant):
\be
e^{-q_n}={|\tilde{g}_n|\over 2\mu^2_n} \biggl[ e^{\mu_n (\tau-\tau_n)}
+ e^{-\mu_n(\tau-\tau_n)} +2\ve_n \biggr] ,
\label{eq:21}
\ee
where $\ve_n \equiv -|\tilde{g}_n| /\tilde{g}_n$,
$\mu_n$ and $\tau_n$ are the integration constants ($\mu_n^2$ and $\tau_n$ are real).
The constraint can be shown to be $\sum_n \mu_n^2 /\lambda_n = 0$,
and its solution is trivial. The space of the solutions is thus defined by the
$(2N-2)$ dimensional moduli space (one of the $\tau_n$ may be fixed).
One can show that the solutions have at most 2 horizons\footnote{
To prove this one should analyze the behavior of $q_n$ for $|\tau| \lim \infty$
and for $|\tau - \tau_n | \lim 0$ (if $\ve_n < 0$). The horizons appear when
$F \lim -\infty$ while $\phi$ and $\psi_n$ for $n \geq 2$ tend to finite limits.
This is possible for $|\tau | \lim \infty$ if and only if $\mu_n = \mu$.
When $F \rightarrow F_0$ and $\phi \rightarrow \infty$ we have the flat space
limit, e.g. exterior of the black hole. The singularities
in general appear for $|\tau - \tau_n | \lim 0$ if $\ve_n < 0$.}
and the space
of the solutions with horizons has dimension $(N-1)$. There exist integrable models
having solutions with horizons and no singularities but their relation to the high
dimensional world is at the moment not clear.

Note that the solution (\ref{eq:21}) is written in a rather unusual coordinate system.
One may write a more standard representation remembering that the dilaton $\phi$
is related to the coordinate $r$ (see \ref{eq:4})). This may be useful for a geometric
 analysis of some simple solutions (e.g. Schwarzschild or Reissner - Nordstr\o m)
 but in general
the standard representation is very inconvenient for analyzing the solutions of the
$N$-Liouville theory.

\section{Discussion and outlook}
The explicitly analytically integrable models presented here
may be of interest for different applications.
The most obvious one is to use them to construct first approximations to generally
non integrable theories describing black holes and cosmologies. Realistic theories
of these objects are usually not integrable (even in dimension 0+1). Having explicit
general solutions of the zeroth approximation in terms of elementary functions
it is not difficult to construct different sorts of (classical) perturbation theories.

For example, spherically symmetric static black holes non minimally coupled to
scalar fields are described by the integrable 0+1 dimensional $N$-Liouville
model. However, the corresponding 1+1 dimensional theory is not integrable
because the scalar coupling potentials $Z_n$ are not constant
(see eq.(\ref{eq:5})). To obtain approximate analytic solutions of the 1+1
dimensional theory one may try to approximate $Z_n$ by properly chosen constants.

It may be useful to combine this approach with the recently proposed analytic perturbation
theory allowing to find solutions near horizons
for the most general non integrable 0+1 dilaton gravity theories \cite{atfm}.
The detailed description of the $N$-Liouville (and of the Toda type) theories,
as well as applications to black holes and cosmology, will be given elsewhere.
The Toda type theories were earlier introduced by direct reductions of higher dimensional
theories to cosmological models
(see e.g.  \cite{Iv}, \cite{Fre} and references therein).

A.T.F. is grateful to V.D.A. for a warm hospitality
at the University of Turin, to P.~Fre, O.~Lechtenfeld, D.~Luest, D.~Maison and A.~Sorin
for interest in this work and useful remarks, to D.~Luest for hospitality at the Humboldt
University in Berlin, to G.Altarelly for hospitality at CERN, and to P.Sorba for
hospitality at LAPP in Annecy. This work was partly supported by RFBR grant 03-01-00781.

\section{Appendix}

\subsection{Reduction of the Curvature}

We usually consider the block diagonal dimensional reduction
\be
ds^2 = g_{ij} dx^i dx^j + h_{mn} dx^m dx^n ,
\label{a1}
\ee
where the metric depends only on the coordinates of the first subspace, $x^i$.

The Ricci curvature scalar for this metric then can be written as
\ba
R = R[g] + R[h] -{2 \over \sqrt{h}} \nabla^m \nabla_m \sqrt{h} +
\quart g^{ij} \p_i h^{mn} \p_j h_{mn} + \nonumber \\
\quart g^{ij} (h^{mn} \p_i h_{mn})
(h^{pq} \p_j h_{pq})  \, .
\label{a2}
\ea
Using this expression, partial integrations, and the Weyl transformations
one may easily derive the reductions presented in the main text.
If the second subspace is a $d-2$ dimensional
sphere of radius $e^{\mu}$ then
$$R[h] = e^{-2\mu} k(d-2)(d-3) ,$$
where $k=1$ for
the sphere ($k=-1$ for the pseudo sphere and $k=0$ for the flat space; these
objects appear in the cosmological reductions (\ref{eq:5a})).

To help the reader in keeping  trace of relations between dimensions $d$, 1+1, 1+0 and 0+1 we
also write here a simple expression for the curvature in dimension 1+1. We take the diagonal
metric
\be
ds^2 = -e^{2\nu}dt^2 + e^{2\mu}dr^2 .
\label{a4}
\ee
The Ricci scalar $R$ for this metric is
\be
R = 2 e^{-2\nu} (\ddot{\mu} + {\dot{\mu}}^2 - \dot{\mu} \dot{\nu} ) -
2 e^{-2\mu} (\nu^{\prime\prime} + {\nu^{\prime}}^2 - \nu^{\prime} \mu^{\prime}) .
\label{a5}
\ee
For this metric, one may also need the expression for $\nabla^2 \phi$,
where $\phi$ is an arbitrary scalar field:
\be
\nabla^2 \equiv \nabla^m \nabla_m \phi = - e^{-2\nu} \bigl( \ddot{\phi} +(\dot{\mu} -
\dot{\nu}) \dot{\phi}) + e^{-2\mu} ({\phi}^{\prime\prime} +
({\nu}^{\prime} - {\mu}^{\prime}) {\phi}^{\prime} \bigr) .
\label{a6}
\ee

All these expressions simplify in the $(u,v)$ coordinates that can be obtained by
taking $\nu = \mu$ and introducing the light cone coordinates, which is always
possible for the 1+1 dimensional metric (having the Minkowski signature).
Denoting $e^{2\nu} = e^{2\mu} = f$ we have
\be
ds^2 = f(dr^2 - dt^2), \,\,\,\,\,\,  R = {1 \over f} (\p_t^2 - \p_r^2) \ln{|f|}
\label{a7}
\ee
At this point one may introduce the $(u,v)$ metric which drastically simplifies the
equations of motion and all computations. One may, for example, write $t = u + v$
and $r = u - v$ and the metric will have the standard form
\be
ds^2=-4\, f(u,v) \, du \, dv \,.
\label{a8}
\ee

However, had we chosen $r = u + v$ and $t = u -v$, what may look more natural in
considering static solutions, the sign in (\ref{a8}) would change. Moreover, there
is a residual symmetry in the $(u,v)$ coordinates, namely, $u \rightarrow a(u)$,
$v \rightarrow b(v)$. Under this transformation (\ref{a8}) transforms as
\be
ds^2=-4\, f(a(u),b(v)) \, a'(u) \, b'(v) \, du \, dv \, = -4\, f(a,b) \, da \, db .
\label{a9}
\ee
Thus the metric in the coordinates $(a,b)$ is the same as in the $(u,v)$ coordinates.
Also the curvature and equations of motion remain invariant.

This freedom is useful for many reasons. For example, suppose we have found
a solution of the equation of motion, for which the metric $f$ and the dilaton
$\varphi$ depend only on $uv$. Then, choosing $a = \ln{u}$, $b = \ln{v}$, we
may go to coordinates $(a,b)$ in which the metric function and the dilaton
depend on $a + b$ (this may be interpreted as $r$ or as $t$)\footnote{
This means that the 1+1 dimensional metric is effectively one dimensional. If
it originated from the higher dimensional spherically symmetric metric (\ref{eq:5a}),
this also should be effectively one dimensional. This, however, does not mean that
the whole theory reduces to one dimension, because the scalar matter fields
may still depend on two variables (see e.g. \cite{FI}).}.
More generally, the solutions of integrable models may usually be written in terms
of massless free fields $\chi_n$ which are solutions of the D'Alembert equation and thus
may be written as a sum of left moving fields $a_n(u)$ and right moving ones $b_n(v)$,
$\chi_n = a_n + b_n$. If all $\chi_n$ are equal, i.e. $\chi_n = a(u) + b(v)$, the theory
reduces to one dimension\footnote{
It is a good exercise to find a free field representation for the $N$-Liouville theory
and to reduce it to one dimension by using this idea.}.
In the same way one may dimensionally reduce the general, non integrable models.
We describe the simplest approach using the light cone coordinates. The more standard
approach uses coordinates $r$ and $t$. It is more cumbersome but may be of use
for interpreting the low dimensional solutions as solutions of higher dimensional theories.

\subsection{Reduction of the Equations}

First, let us write the equations of motion in the light cone ($u,v$) coordinates
(their derivation from the Lagrangian (\ref{eq:7}) is a good exercise for the reader).
To simplify the formulas we keep only one scalar field:
\be
{\cal L} = \sqrt{-g}\, \biggl( \f R + V(\f,\psi) +Z(\nabla \psi)^2 \biggr).
\label{a11}
\ee
In the $(u,v)$ coordinates they are
\ba
&& \p_u \p_v \f+f\, V(\f,\psi)=0, \label{F.15} \\
&&\p_v (Z \p_u \psi) +\p_u (Z \p_v \psi) + f V_{\psi}(\f,\psi)=
Z_{\psi} \, \p_u \psi \, \p_v \psi,
\label{F.16} \\
&& f \p_i ({ {\p_i \f} \over f }) \, = Z (\p_i \psi)^2, \quad i=u,v, \label{F.17} \\
&&\p_u\p_v\ln |f| + f V_{\f}(\f,\psi) = Z_{\f} \,\p_u \psi \, \p_v\psi,
\label{F.18}
 \ea
 where $V_{\f}= \p_{\f} V$, $V_{\psi}= \p_{\psi} V$,
 $Z_{\f}= \p_{\f} Z$, $Z_{\psi} =\p_{\psi} Z$.
  Equations (\ref{F.15}) $-$ (\ref{F.18}) are not independent. Actually,
  (\ref{F.18}) follows from equations
(\ref{F.15}) $-$  (\ref{F.17}). Alternatively, if  (\ref{F.15}),
(\ref{F.17}), (\ref{F.18}) are satisfied,
   (\ref{F.16}) is satisfied.

The most important equations are the constraint equations (\ref{F.16}).
A general formulation of the dimensional reduction is suggested be the
following simple observation. Consider the solutions with constant scalar
field $\psi \equiv \psi_0$ (the `vacuum' solution). This solution exists if
and only if $V_{\psi}(\varphi,\psi_0) = 0$, see eq.(\ref{F.16}).
Then the constraints can
be solved because their right-hand sides are identically zero. It is not
difficult to prove (this is a simple exercise) that there exist chiral
fields $a(u)$ and $b(v)$ such that
\bdm
\varphi (u,v) \equiv \varphi (\tau), \,\,\,\,\, {\rm and} \,\,\,\,\,
f(u,v) \equiv \varphi^{\prime} (\tau) \, a'(u) \, b'(v) ,
\edm
where the primes denote derivatives with respect to the corresponding argument.
Using this result it is easy to prove that eq.(\ref{F.15}) has the integral
\bdm
\varphi^{\prime} + N(\varphi) = M ,
\edm
where $N(\varphi)$ is defined by the equation $N^{\prime}(\f) = V(\f , \psi_0)$
and $M$ is the integral of motion which for the black hole solutions is proportional
to the mass of the black hole. The horizon, defined as a zero of the metric
$h(\tau) = M - N(\varphi)$, exists because the equation $M = N(\varphi)$ has at least
one solution in some interval of values of $M$. We see that the equations of motion
in the considered case are actually dimensionally reduced. Their solutions can be
interpreted as black holes (including the Schwarzschild, the Reissner Nordstr{\o}m and
other known black hole solutions in any dimension) or as cosmological models (including
the Friedmann - Robertson - Walker cosmology and its generalizations).

The more general static solutions with horizons and more general cosmologies
are not described by the scalar vacuum solutions. They are not the general solutions
of the 1+1 dimensional equations and are derived as solutions of differently chosen
one dimensional sectors of the 1+1 dimensional theory. Here we introduce the simplest
dimensional reduction. As we work in the light cone coordinates, the interpretation
and comparison to the standard considerations requires reintroducing the $(t,r)$
coordinates as it can be done simply and generally.

Taking into account the lesson of the scalar vacuum solutions, we introduce the dimensional
reduction by supposing that the scalar fields and the dilaton depend on one free field
$\tau$ (after dimensional reduction it is interpreted as the space or the time coordinate):
\be
\f=\f(\tau), \quad \psi=\psi(\tau), \qquad \tau=a(u)+b(v).
\label{a12}
\ee
Then, it follows from eq.(\ref{F.15}) that the metric should have the form
\be
f(u,v) =\ve \, h(a+b)\, a'(u) \, b'(v),
\label{a13}
\ee
where $\ve$ is introduced in order to have the same type of metric for the
0+1 and 1+0 cases:
\ba
ds^2=h(\tau) \bigl(dr^2 -dt^2\bigr)
\label{a13a}
\ea

Usually one defines the $r$ and $t$ coordinates in terms of $u$ and $v$.
More generally, we may define them in terms of $a(u)$ and $b(v)$. Defining
\be
 \tau\equiv a+b, \,\,\,\,\, \bar \tau\equiv a-b , \label{a14}
\ee
we have from eq.(\ref{a13})
\be
 ds^2--4f(u,v) \, du \, dv=-4\ve h \, da \, db=-\ve h \bigl(d\tau^2-d \bar
\tau^2\bigr) \, , \label{a15}
\ee
and thus both reduced metrics may be written as (\ref{a13a}) by choosing
 $\tau=r$, $\ve=-1$ or $\tau=t$ and $\ve=+1$.

The reduced equation of motion for the dilaton and for the scalar field,
\bdm
\p_{\tau}^2 \f +\ve hV=0 \, ,  \qquad
2 \p_{\tau} \bigl( Z\p_{\tau}\psi\bigr) +\ve h V_{\psi} = Z_{\psi}
\bigl(\p_{\tau}\psi\bigr)^2 \, ,
\edm
depend on $\varepsilon$ while the constraints are the same for both reductions
and give just one reduced constraint,
\bdm \p_{\tau}^2 \f - \p_{\tau}\f \, \p_{\tau} \ln|h| =
Z(\p_{\tau}\psi)^2 \, ,
\edm
that is equivalent (in the standard terminology) to the energy constraint.

Thus we have {\it the rule} for the reduction of the equations of motion:
using the equations in the light cone gauge, derive the equations for
$\f(\tau)$, $\psi(\tau)$, $\ve h(\tau)$ and then take
$\tau=r$ and $\ve=-1$  or $\tau=t$  and $\ve=+1$.

One may avoid writing the 1+1 dimensional equations of motion by directly
reducing the Lagrangian (\ref{a11}). To do this one may start from the 1+1
dimensional Lagrangian  in the general diagonal metric,
\bdm
ds^2= -h_0 dt^2+h_1 dr^2 \, ,
\edm
and derive  the Lagrangian for the 0+1 reduction ($h_0(r), \, h_1(r)$)
and the 1+0 reduction ($h_0(t),\,\, h_1(t)$) separately.
However there is a simpler and more direct way which allows to write the
correct equations without calculations.

First, take the gauge fixed Lagrangian in the $(u,v)$ metric,
\be \label{a20}
{\cal L}=\f \, \p_u\p_v F +f V-Z \, \p_u \psi \, \p_v\psi
\ee
where $F=\ln |f|$. Due to the residual covariance with respect to the
transformation $u\ra a(u)$, $v\ra b(v)$, we may equivalently write
\be
{\cal L}=\f \p_a \p_b F +\ve\, h V - Z \, \p_a \psi \, \p_b \psi .
\label{a21}
\ee
Then, substituting in this Lagrangian (\ref{a12}) and (\ref{a13}) we obtain
\bdm
{\cal L} =\f \ddot F  -Z\dot \psi ^2 + \ve h V ,
\edm
where the dot denotes $\tau$- differentiation.   This Lagrangian is equivalent to
\be   \label{a22}
{\cal L}= -\dot \f \dot F -Z \dot \psi^2 +\ve h V ,
\ee
and it gives the correct reduced equations of motion.

 Let us restore the lost constraint (the gauge fixed Lagrangian (\ref{a21}) does not
give the constraints). To do this we recall that the constraint is just
${\cal H}=0$, where ${\cal H}$ is the Hamiltonian correspondent to the Lagrangian.
It is evident that
\be
{\cal H} =-\dot \f \dot F -Z\dot \psi -\ve h V.
\label{a23}
\ee
Now it is easy to guess that the correct Lagrangian giving the equations
of motion and the constraint ${{\cal H} = 0}$ is simply
\be
{\cal L} = - {1\over l(\tau) } \biggl( \dot\f \dot F + Z\dot \psi^2 \biggr)
+ l(\tau) \ve h V.
\label{a24}
\ee
In order to obtain from here the 0+1 theory we simply take $\tau=r$ and
$\ve=-1$. The 1+0 theory can be written taking $\tau=t$ and $\ve=+1$.

Finally, let us write an example of cosmological reduction directly from
a higher dimensional theory. We take the $d$ dimensional metric (\ref{eq:5a}),
suppose that the scalar functions depend on one variable $t$ (the gauge field
is reduced differently, see e.g. \cite{St}. Then using eq.(\ref{a2}) with
the one dimensional metric $g$ and the $d-1$ dimensional metric $h$ we can find
for example the reduced action for the $d$ dimensional Lagrangian (\ref{eq:3b}).
We write here only the reduced curvature part (the derivation of the other terms
is obvious):
\ba
S = \int d^d x \, \sqrt{-g} \, \sqrt{h} \, R^{(d)} = \int dt \, e^{\nu} \, e^{\mu (d-1)}
\bigl[ \, k(d-1)(d-2) e^{-2\mu} - \nonumber \\
- (d-1)(d-2) e^{-2\nu} \, {\dot{\mu}}^2 \, \bigr] . \,\,\,\,
\label{a25}
\ea

\subsection{Nonlinear coupling of gauge fields}

Suppose that, in place of the standard Abelian gauge field term,
$X(\f,\psi) F^2 $, the Lagrangian contains a more general coupling of
the gauge field $F_{ij}=\p_i A_j -\p_j A_i$
to dilaton and scalar fields, ${\cal F}(\f, \psi; F^2)$ (for example, one
may consider the Born - Infeld type terms). The equation of
motion for the gauge field is
\be     \label{F.01}
\p_i \biggl( \sqrt{-g} \, {\p {\cal F} \over \p(\p_i A_j)} \biggr)
=0,
\ee
where
\bdm    
{\p {\cal F} \over \p (\p_i A_j)} = 4 F^{ij} {\p{\cal F} \over \p F^2},
\edm
reduces in 1+1 dimension to the conservation law
 \be  \label{F.02}
 \sqrt{-g} \, F^{ij} {\p {\cal F} \over \p F^2} = \varepsilon ^{ij}
 \lambda Q
 \ee
 where $\ve^{ij}=-\ve^{ji}$, $\ve^{01}=1$, $\lambda$ is a constant to be
 defined later and $Q$ is a conserved charge.
Using this equation
we can express $F_{ij}$ in terms of $F^2$:
\be \label{F.03}
F^{ij}={\ve^{ij} \lambda Q\over \sqrt{-g}} \biggl( {\p {\cal F}\over \p
F^2} \biggr)^{-1}.
\ee
From this it is easy to obtain the equation for $F^2$ (recall that
$2g=\ve^{ij} \ve^{lk} g_{il} g_{jk}$):
\be  \label{F.04}
F^2  = -2 \lambda^2 \,Q^2 \biggl( {\partial {\cal F}
\over \partial F^2} \biggr)^{-2}.
\ee
This allows (in principle) to write $F^2$ (and  $F_{ij}$) in terms of
$\f,\,\psi,\, Q$. Let us denote the solution as
$\bar F^2\equiv \bar F^2(\f,\psi;Q)$ (or simply $ \bar F^2)$.
Now we can write $\bar F^{ij} $ in terms of $\bar F^2(\f,\psi;Q)$.
Equation (\ref{F.04}) gives
\be
{\p \bar {\cal F} \over \p F^2} \equiv {\p {\cal F} \over \p F^2}_{|_{F^2
= \bar F^2}} = \epsilon {\sqrt{2} \lambda Q} /
 {\sqrt{-\bar F^2}} ,
\label{F.05}
\ee
where $\bar {\cal F} \equiv {\cal F}(\f, \psi; \bar F^2)$,
$\epsilon = {\rm sign} [ \, {\p {\cal F} / \p F^2} ]$ and
$\sqrt{-F^2} > 0$\footnote{
For small values of $F^2$ we have usually $\epsilon < 0$. }.

 Then from equations (\ref{F.03}) and (\ref{F.05}) we get
 \be  \label{F.06}
 \bar F^{ij}={\ve^{ij} \lambda Q \over \sqrt{-g}} \, \biggl( {\p \bar {\cal F}
 \over \p F^2} \biggr)^{-1}=\ve^{ij}\,\epsilon\, \, {\sqrt{ \bar F^2 / 2g }} .
 \ee
Now we can exclude the gauge fields
from the equations of motion. In order to do
this, let us find an effective potential ${\cal F}_{eff}$ depending only
on $\f,\psi$ and $ \bar F^2(\f,\psi;Q)$.
To simplify the computation we go to the $(u,v)$ coordinates in which
\be \label{F.07}
F^{uv}={1\over 4 f^2} F_{uv}, \qquad F^2=-{1\over 2f^2} F_{uv}^2.
\ee
Note that in order to obtain the complete equations of motion including the
constraints we should use the $(u,v)$ coordinates {\it after} computing the
variations of ${\cal L}$ in the diagonal metric  coefficients $g_{ii}$.
Only then we may set $g_{ii}=0$ and $g_{uv}=g_{vu}=-2f$.
Note also that the other variations (in $\f, \psi$) can be derived from
the $(u,v)$ reduced (gauge fixed) Lagrangian
\be   \label{F.08}
{\cal L}= \f \p_u \p_v  \ln |f| +fV(\f,\psi) -Z\p_u   \psi
\p_v \psi + f{\cal F}(\f,\psi;F^2).
\ee

In order to  find the effective potential ${\cal F}_{eff}$
 we derive the expression for the variation $\delta (f {\cal F})$ with respect to $f$.
We need not calculate the variation with respect to $g_{ii}$ because
 \bdm
 {\delta F^2 \over \delta g^{ii}} =2 g^{jk} F_{ij} F_{ik}\equiv 0 \quad
 {\rm when} \quad  g_{jj}=0
 \edm
 (this also means that the constraints are insensitive to the ${\cal F}^2(\f,
 \psi,F^2)$ term  when we use the $(u,v)$ coordinates).

Now we have (we don't set yet $F^2=\bar F^2$)
\be  \label{F.09}
\delta_f\bigl(f {\cal F}\bigr) = {\cal F} \,\delta f + f {\p {\cal F}\over
\p F^2} \cdot {\p F^2\over \p f} \cdot \delta f =
\bigl({\cal F} \,-\, 2F^2 {\p {\cal F}\over \p F^2}\bigr)\, \delta f
 \ee
because
\bdm
{\p F^2\over \p f} = {\p \over \p f} \biggl(-{1\over 4f^2} F_{uv}^2 \biggr)
=-{2\over f} F^2.
\edm
Using in (\ref{F.09}) the relations (\ref{F.05}) and (\ref{F.06}) we
obtain
\be \label{F.10}
\delta_f(f {\cal F} ) |_{F^2=\bar F^2}=\delta f\biggl[ {\cal F}(\f,\psi,\bar F^2)
+2\sqrt{2} \lambda  Q \epsilon \sqrt{-\bar F^2}\,\biggr].
\ee

The right hand side of (\ref{F.10}) produces the effective potential we are
looking for,
\be  \label{F.11}
{\cal F}_{eff}(\f,\psi,\bar F^2)\equiv  {\cal F}(\f,\psi; \bar F^2) \,
+ \,2\sqrt{2} \,\lambda Q \epsilon \sqrt{-\bar F^2}.
\ee
To prove this it is sufficient to differentiate the derivatives of
${\cal F}_{eff} $ with respect to $\bar F^2$, to $\f$ and to $\psi$. From
\bdm
{\p {\cal F}_{eff} \over \p \bar F^2} = {\p {\cal F}\over \p \bar F^2} -
\epsilon { \sqrt{2} \lambda Q}  / {\sqrt{-\bar F^2} }
\edm
we see that the main equation defining $\bar F^2$ will be reproduced if we
set $\p {\cal F}_{eff}/\p \bar F^2 =0$  (as we should do, because now
we consider $\bar F^2$ as a new variable).  If we now express ${\cal F}_{eff}$ in
terms of $\f,\psi,Q$, i.e.
\bdm
{\cal F}_{eff}\bigl(\f,\psi;\bar F^2(\f,\psi,Q)\bigr)\equiv
{\cal F}_{eff}(\f,\psi ; Q),
\edm
we find that
\bdm
{d {\cal F}_{eff} \over d\f}={\p {\cal F}\over \p \f}  +
\biggl( {\p {\cal F} \over \p \bar F^2}  -
\epsilon \sqrt{2} {\lambda Q / \sqrt{- \bar F^2}} \biggr) {\p \bar F^2 \over \p \f }
=  {\p {\cal F}\over \p \f}
 \edm
 due to (\ref{F.05}). Analogously,
${d {\cal F}_{eff} / d\psi}={\p {\cal F} / \p \psi}$.
This means that the $\f,\psi,f$ equations of motion  for the effective
Lagrangian ${\cal L}_{eff}$ (in which ${\cal F}$ is replaced by ${\cal F}_{eff}$)
coincide with the $\f,\psi,f$ equations of motion
for the Lagrangian ${\cal L}$ and give the correct expression for $\bar
F^2$ and $\bar F_{ij}$. (Note that we did not use the equations of motion for
$\f,\psi,f$ in our derivation of ${\cal F}_{eff}$.)

Thus we can  include ${\cal F}_{eff}$ into the effective potential, i.e. define
\be  \label{F.14}
V_{eff}(\f, \psi;Q) =V(\f,\psi) \,+\, {\cal F}_{eff}(\f,\psi;Q)
\ee
and forget about the fields $F_{ij}$, $A_i$ that can be derived from eq.
(\ref{F.06}) if needed.

It is not difficult to understand that ${\cal F}$ may depend on
any number of  fields $\psi$ and any number of abelian gauge fields
$F_n^2$. Thus in general the effective potential  will be given by
\bdm
{\cal F}_{eff} = {\cal F} (\f,\psi_m;\bar F_n^2) + 2\sqrt{2}\, \lambda
\, \sum_n Q_n \epsilon_n \sqrt{-\bar F_n^2}
\edm
where  the quantities $\bar F_n^2$ are solutions of the equations

\bdm
 F_n^2\, =\, -\,2\lambda^2 Q_n^2\,\biggl( {\p {\cal F} \over \p
F_n^2} \biggr)^{-2}   \,\,\, {\rm and  } \,\,\, \epsilon_n \equiv {\rm sign}
\biggl[ {\p{\cal F} \over \p F_n^2} \biggr] .
\edm

Now consider a simple example. Let us take the Born - Infeld expression
for ${\cal F}$
\bdm
{\cal F}=\alpha\sqrt{\beta^2 \pm F^2}
\edm
where $\alpha$ and $\beta$ are functions of $\f,\psi$ ($\beta>0$).  Applying our
formulae we get the following effective potential:
\bdm
{\cal F}_{eff}=\beta {\alpha \over |\alpha|}
\sqrt{\alpha^2 \pm \bar Q^2}\qquad {\rm where} \quad \bar Q^2
\equiv 8\lambda^2 Q^2.
\edm
Now, if $\alpha$ is a constant, $\alpha=\alpha_0$, and $\beta =g^2
\exp(2\mu \f + 2\nu \psi)$ we obtain an exponential effective
potential:
\bdm
{\cal F}_{eff} =g e^{\mu \f +\nu \psi} {\alpha \over |\alpha|}
\sqrt{\alpha_0^2 \pm \bar Q^2}.
\edm

Consider now the case $F^2<<\beta^2$. Then
\bdm
{\cal F} =\alpha \,\beta\, \bigl( 1 \pm {F^2\over 2\beta^2} +...\bigr)
        =\alpha \beta \pm {F^2\over 2\beta/\alpha} +...
 \edm
Using our formula we have the well known result
\bdm
{\cal F}_{eff}=\alpha \beta \pm 2 \lambda^2 Q^2 \bigl({2\beta\over \alpha}\bigr)
+...
\edm

\end{document}